\nonstopmode
\documentstyle[11pt,psfig,epsf]{article}
\def\reference{\parskip 0pt\par\noindent\hangindent 0.5 truecm}

\def\lesssim{\mathrel{\hbox{\rlap{\hbox{\lower4pt\hbox{$\sim$}}}\hbox{$<$}}}}
\def\gtrsim{\mathrel{\hbox{\rlap{\hbox{\lower4pt\hbox{$\sim$}}}\hbox{$>$}}}}
\newcommand{\et}{et al.}

\begin{document}
%
%
\title{Circular Polarisation in AGN}
%

\author{Jean-Pierre Macquart\footnote{Present address: Kapetyn Astronomical Institute, University of Groningen, The Netherlands}
} 

\date{}
\maketitle

{\center
School of Physics, University of Sydney, NSW 2006, Australia
\\jpm@physics.usyd.edu.au\\[3mm]
}

\begin{abstract}
We discuss the constraints that recent observations place on
circular polarisation in AGN. In many sources the circular
polarisation is variable on short timescales, indicating that it
originates in compact regions of the sources.  The best prospects
for gleaning further information about circular polarisation are
high resolution VLBI and scintillation `imaging' which probe
source structure on milliarcsecond and microarcsecond scales
respectively.
\end{abstract}



\medskip


\section{Introduction}
The mean degree of circular polarisation (CP) in AGN is typically
less than 1\%, but its origin is not understood.  Although the
linear polarisation observed in AGN is accepted as evidence of
synchrotron emission, the physical significance --- hence the
correct interpretation
--- of the CP in AGN is not clear.  This is largely because, until now,
the observations were of insufficient quality to constrain the properties
of the CP.

New data from the VLBA and the Australia Telescope Compact Array
(ATCA) have enabled a reassessment of the origin of CP. The high
angular resolution of the VLBA gives it a decisive advantage over
most telescopes in circular polarimetry.  Since the magnitude of
the CP is usually small and occurs primarily in compact regions,
high resolution is necessary to mitigate the effects of beam
depolarisation.  The VLBA has also been able to, for the first
time, determine the location of a circularly polarised region in
an AGN relative to its core and jets.

Paradoxically, the ATCA, with its modest angular resolution, has
made equally important contributions to our understanding of CP.
Its advantage is the demonstrated ability to measure the degree of
CP in sources to better than 0.01\%.  It is particularly useful
when the angular resolution of the instrument is irrelevant
because there is some other means of obtaining high angular
resolution information about the source, usually from source
variability.  Its ability to do high precision, short timescale
monitoring of intraday variable (IDV) radio sources, coupled with
an understanding of interstellar scintillation, allows it to
achieve microarcsecond resolution polarimetry.

We outline in Section 2 the various possible origins of the CP and
in Section 3 the constraints placed upon these mechanisms by the
observations. Various other sources that exhibit CP are discussed
in Section 4.  The conclusions are presented in Section 5.

\section{Origins of Circular Polarisation}
\subsection{Synchrotron Emission}
There is a small degree of CP associated with synchrotron
emission. For an electron with Lorentz factor $\gamma$ gyrating
around a field line at an angle $\theta$ to the line of sight the
degree of CP observed is (Legg \& Westfold 1968)
\begin{eqnarray}
    m_c = \frac{\cot \theta}{3} \left( \frac{\nu}{3 \nu_H} \right)^{1/2}
     \approx \cot \theta/\gamma  \label{SynchForm}
\end{eqnarray}
in an optically thin source, where $\nu_H = e B \sin \theta/m_e$ is the
electron gyrofrequency.  The degree of CP for an optically thick source
remains similar; it is modified by a term of order unity dependent on the
distribution of relativistic electrons, $N(\epsilon) \propto \epsilon^{-2
\alpha -1}$ (Melrose 1971).  In a homogeneous source the sense of the CP
is expected to reverse as the source becomes optically thick.

Equation (\ref{SynchForm}) implies that high CP is associated with
emission from particles with low Lorentz factors, $\gamma$, or
emission nearly parallel to the magnetic field (i.e.\  small
$\theta$).  However, in both instances the power expected from
synchrotron radiation is small. The brightness temperature of
particles radiating with Lorentz factor $\gamma$ is limited by
self-absorption to $T_B \lesssim (\gamma-1) m_e c^2 /k \approx 0.6
\times 10^{10}\, (\gamma-1)$~K.  However, bulk relativistic motion
of the emitting region toward the observer can Doppler boost the
radiation, permitting higher degrees of CP for a given observed
brightness temperature.

The emission intensity in synchrotron radiation is proportional to
$(\sin \theta)^{\alpha + 1}$.  This falls to zero for emission
nearly parallel to the magnetic field lines (i.e.\ as $\theta
\rightarrow 0$). Although a high degree of CP is expected from
radiation parallel to the magnetic field, it is associated with
low power emission.  Furthermore, regions with small $\theta$
would be confined to a small volume of the source in any plausible
model for the distribution of the magnetic field.  The importance
of magnetic field non-uniformity is implied by the fact that the
degree of linear polarisation observed in AGN is at least an order
of magnitude lower than the $\sim 70$\% expected due to
synchrotron radiation.  This suggests that the level of CP
expected due to synchrotron radiation is at least an order of
magnitude lower than that expected from equation
(\ref{SynchForm}).

The low levels of CP expected from this emission mechanism and the
fact that its $m_c \propto \nu^{-1/2}$ dependence is generally not
observed (see references in Saikia \& Salter 1998) suggest that
synchrotron emission is not responsible for most of the CP
observed in extragalactic sources at centimetre wavelengths.

\subsection{Cyclotron Emission}
Cyclotron emission refers to radiation from non-relativistic
electrons (i.e.\ $\gamma -1 \ll 1$).  If the electron orbit is
edge-on to the line of sight then 100\% linearly polarised
radiation is observed and, if face-on, the radiation is fully
circularly polarised.  Elliptically polarised radiation is
observed for intermediate orientations.

Cyclotron emission occurs at frequencies that are integer
multiples of the cyclotron frequency, $\nu_H$.  The power of the
emission at the harmonics of the cyclotron frequency is
proportional to $\sin \theta$, and thus falls to zero as $\theta
\rightarrow 0$.  The radiative power in these harmonics therefore
falls to zero as the emission becomes more circularly polarised.
The radiative power is non-zero for $\theta=0$ only for the
emission at the fundamental frequency, and this is the only
radiation which may be, in principle, fully circularly polarised.

The brightness temperature of the emission in the rest frame of the
emitting particles is low: $T_B \sim 0.6 \times 10^{10}$~K.  Unless the
cyclotron emission is highly Doppler boosted any CP associated with it is
easily diluted by emission from brighter, synchrotron emitting regions.

\subsection{Coherent Emission Mechanisms}

Coherent emission processes have the potential to account for CP
in AGN. The polarisation of pulsar radiation suggests that this is
plausible. Pulsar radiation must be due to coherent emission,
since the brightness temperature of the radiation is in the range
$T_B \approx 10^{22}-10^{26}$~K. The emission is also observed to
be highly polarised (e.g.\ Manchester, Taylor, \& Huguenin 1975),
however it is possible that the CP in pulsars is not intrinsic to
the radiation mechanism at all, but instead arises due to
propagation effects in the pulsar magnetosphere.


It is difficult to make specific comments about the polarisation
properties of any coherent emission that may occur in AGN since
(i) there is no preferred mechanism for the coherent emission,
(ii) the polarisation properties of  most of the various
mechanisms have not been considered in detail, and (iii) it is
unknown whether the physical conditions necessary for the coherent
emission would also lead to depolarisation of the escaping
radiation.  This last point is relevant because any coherent
emission observed in AGN is likely to emanate from many localised
patches emitting simultaneously (e.g.\ Melrose 1991; Benford
1992), which would lead to averaging out of the polarisation
properties of each individual patch.  Nonetheless, some specific
mechanisms predict no CP (Windsor \& Kellogg 1974; Benford 1984).

Coherent mechanisms are not required to explain high brightness
temperature emission in IDV sources (see Melrose, these
proceedings), and they do not seem to be explicitly required to
explain the few per cent levels of CP seen in these sources
either. Furthermore, if these processes do operate in AGN it is
not obvious that they would dominate incoherent synchrotron
emission or propagation induced polarisation effects.


\subsection{Faraday Conversion in a Relativistic Plasma}

Propagation can convert CP from linear polarisation in a process
analogous to Faraday rotation, called generalised Faraday rotation
or Faraday repolarisation (Sazonov 1969; Pacholczyk 1973; Jones \&
O'Dell 1977).

The transfer of polarised radiation in a birefringent plasma may
be understood in terms of the Poincar\'e sphere (Figure~1).  A
magnetised homogeneous plasma has two mutually orthogonal natural
wave modes, with some difference in wavenumber between the two
modes, $\Delta k \neq 0$. The wave modes of a given birefringent
medium define a diagonal through the origin and opposite ends on
the surface of the Poincar\'e sphere. Propagation through the
plasma induces a phase difference $L \Delta k$ between the two
orthogonally polarised wavefields which results in a rotation of
the polarisation vector around the Poincar\'e sphere at constant
latitude around the axis defined by the natural modes.

In a magnetised relativistic plasma the natural modes of the
medium are elliptically polarised.  Propagation of radiation
through such a plasma converts linearly polarised radiation into
circularly polarised radiation as follows (Kennett \& Melrose
1998):
\begin{eqnarray}
V(\nu) &=& U_{0}(\nu) \sin ( \lambda^3 {\rm RRM}), \\
{\rm RRM} &=& 3 \times 10^4 \left( \frac{L}{1\,{\rm pc}} \right)
\left\langle {\cal E}_L \left( \frac{n_r}{1\,{\rm cm}^{-3} } \right)
\left(\frac{B}{1\,{\rm G}} \right)^2 \sin^2 \theta
\right\rangle {\rm rad}\,{\rm m}^{-3}, \label{VUrel}
\end{eqnarray}
where the relativistic rotation measure, RRM, depends upon the
density of relativistic particles, $n_r$, the path length, $L$,
the magnetic field, $B$, and the minimum Lorentz factor of the
pairs, ${\cal E}_L$. This effect operates only when the direction
of the incident linear polarisation is at an oblique angle to the
projection of the magnetic field on the plane orthogonal to the
ray direction.  The axes used to define the Stokes parameters may
be chosen such that synchrotron emission has $Q \neq 0,\, U=0$.
With this choice, the effect occurs only if the incident radiation
has $U_0 \neq 0$, requiring either Faraday rotation or that it
originate from a region of the source where the magnetic field is
in a different direction to that in the region where the
polarisation conversion takes place.

A characteristic of this model is a strong frequency dependence on the
sign of $V$.  If ${\rm RRM}$ is high enough to produce appreciable amounts
of CP at high frequency, this model predicts rapid changes in its
handedness at low frequency. In a relativistic plasma, the position angle
of the radiation defined by the $U-V$ plane rotates $\propto \lambda^3$.

However, the structure of the source can have important
consequences for the frequency dependence of the CP.  The
Blandford-K\"onigl  jet is a particular and relevant instance of
this (Blandford \& K\"onigl 1979). Its largely self-similar
structure can ensure that, at each frequency, the CP is of
constant sign and comparable magnitude. This is because the
emission height of the CP observed along the jet varies as a
function of frequency.

One specific configuration for producing moderate (a few per cent)
CP with consistent sign (see Begelman, these proceedings) involves
a plasma with a weak uniform magnetic field and a highly spatially
variable random magnetic field.  The random component, with high
rotativity per unit length, can generate moderate degrees of CP.
Because this model relies on a highly spatially inhomogeneous
magnetic field, one expects that the CP, as observed on the plane
of the sky, is also highly spatially inhomogeneous. The viability
of this model can be constrained from scintillation observations,
as described in Section 3.

\begin{figure}
\begin{center}
\centerline{\psfig{file=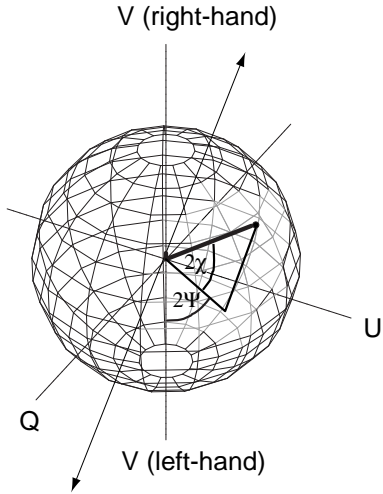,width=50mm}}
\end{center}
\caption[The Poincar\'e sphere]{The Poincar\'e sphere may be used
to represent the polarisation state of radiation.  Elliptically
polarised radiation is mapped from the Stokes parameters ${\cal
P}/I=[Q,U,V]/I$ to the coordinates $[\cos (2 \chi) \cos(2 \Psi),
\cos (2 \chi) \sin(2 \Psi), \sin(2 \chi)]$.  The arrowed diagonal
depicts a case in which natural modes of a medium are elliptical.
Generalised Faraday rotation in such a medium corresponds to
rotation of the polarisation vector at constant latitude about
this axis.}
\end{figure}

\subsection{Scintillation-{\it induced} Circular Polarisation}
Another model that relies upon a highly inhomogeneous medium
involves scintillation-induced CP (Macquart \& Melrose 2000a,b).
This is produced when even unpolarised radiation propagates
through a region with non-uniform $n_e B$ (related to the Faraday
rotation per unit length).  A transverse gradient in this quantity
leads to different ray bending for the right- and left-hand
circularly polarised radiation.  Since the wavefronts are also
rippled due to the phase delays caused by the inhomogeneity, as
occurs in scintillation, transverse displacement of the two
wavefronts can induce CP.  The difference in bending angles
between the two senses of polarisation is small but the radiation
propagates over a large distance, so that the displacement between
the two wavefronts, and hence the CP, can be appreciable.  This
theory may provide an explanation for the variable CP observed at
cm wavelengths in some AGN and in Sgr A*.

There is no preferred handedness for the sense of the CP produced by this
mechanism: although the mean square CP is non-zero, the mean CP, averaged
over a sufficiently long time span, is zero. One expects sign changes in
the CP on a timescale associated with the scintillation pattern.  This is
not observed.  However, this model needs to be extended to remove some of
the constraining assumptions on which it is based.  The most notable of
these is the assumption that the inhomogeneities occur on a thin screen.
The relaxation of this assumption to take into account the extended nature
of the medium would affect both the magnitude of the effect and the
timescale on which sign changes are predicted to occur.

\section{Observational Properties}

In this section we concentrate on the connection between the CP and its
variability.  Rayner (these proceedings) contains a further discussion on
the origins of the CP in AGN.

The CP in AGN was known to be variable almost twenty years ago
(e.g.\ Weiler \& de Pater 1983; Komesaroff \et\ 1984).  In their
monitoring of a sample of AGN at 5~GHz, Komesaroff \et\ (1984)
observed several general features which have been borne out by
more accurate measurements:
\begin{itemize}
\itemsep -2mm
\item Changes in CP occur with only small accompanying
changes in the total intensity.  Interpreting these in terms of an
intrinsic change in the source due to, say, the emergence of a new
component, the authors argued this implied degrees of CP typically
$\sim 5$\% in the polarised subcomponents of these sources. (These
arguments also apply if the variability is due to scintillation.)
\item The timescale of variability in CP is shorter than that of the
linear polarisation, which in turn is shorter than that in the
total intensity.
\item The magnitude of the fractional variability in the CP usually
exceeds that in the linear polarisation, which exceeds that in the
total intensity.
\end{itemize}

Comparison with recent ATCA and VLBA observations shows that many
sources retain the same handedness of CP over decade-long time
scales.  However, early measurements indicated that some sources
do change sign: Komesaroff \et\ (1984) noted sources 2 out of 14
on the $\sim 100$~day timescale that the observations were
sensitive to. The reality of these sign changes has not been
pursued with recent observations.

The CP appears to occur in compact regions.  VLBI observations by
Homan \& Wardle (1999) and Homan, Attridge, \& Wardle (2001)
support the general conclusion that the CP is associated with the
compact regions near the cores of AGN.

It is not clear why some sources exhibit high degrees of CP. There
appears to be no connection between the degree of CP and the
linear polarisation (Rayner, Norris, \& Sault 2000).  VLBI
suggests this may be  due to Faraday depolarisation of the
linearly polarised radiation (e.g.\ Homan et al.\ 2001).  There
is, however, a tentative connection between high CP and IDV AGN.
Many of these sources exhibit strong {\it and} variable CP; these
sources include PKS~1519$-$273 (Macquart \et\ 2000),
PKS~0404$-$385 and PKS~1144$-$379.   In each case the degree of CP
implied by comparing the changes in Stokes $V$ and $I$ is a few
per cent\footnote{The {\it mean} degree of CP in these three
sources ranges from $\approx 0.2 - 0.8$\%.}. The last two sources
exhibit CP variability on shorter timescales than that observed in
the total intensity, suggesting that it emanates from an even more
compact region that does the emission responsible for the bulk of
the unpolarised variability.

Are the degrees of CP inferred in IDV sources atypical?  The
possibility that many other sources have high CP in regions so
compact that they appear beam depolarised even on VLBI scales
remains to be investigated.

\subsection{The Significance of CP Variability}

The prevalence of CP variability in AGN suggests that it occurs
predominantly in compact regions.  One might expect that it {\it
should} vary more rapidly than $I$ because it can also change
sign.  However, this is unlikely to be the correct explanation for
the short timescale. Both Stokes $Q$ and $U$ can also change sign
in the same manner as the CP, but they are not, in general,
observed to vary as quickly.  The rarity of sign reversals in the
CP also argues against this, especially when compared to the large
position angle changes observed in linear polarisation in the same
sources.


The fact that the simultaneous total flux density and polarised
flux density variations in IDV sources are different is often used
as evidence against the viability of scintillation in these
sources.  While scintillation affects all four Stokes parameters
of the propagating radiation identically\footnote{This is strictly
only true if the medium is unmagnetised, but it holds to a good
approximation for scintillations in our Galaxy's ISM.}, the
structure of the scintillating source complicates the issue.

A simple source model comprised of two compact (scintillating)
components illustrates this.  Suppose one component is faint but
highly (say 10\%) circularly polarised, while the other is strong
but unpolarised.  If the angular separation between the sources is
comparable to the angular scale of the scintillations ($\sim
\theta_{\rm F}=r_{\rm F}/D$ for weak scintillations) then the
fluctuations due to each of the components are uncorrelated.
Furthermore, because the fluctuations in CP come from one
component while the majority of the unpolarised flux density
variations originate from the other, the fluctuations in Stokes
$V$ and $I$ show only a small correlation.

The nature of the correlation may change during the course of a
year if the direction of the scintillation velocity changes. Such
changes are possible because changes in Earth's velocity, due to
its orbit about the Sun, can substantially alter the apparent
scintillation velocity. If the two source components lie in a line
parallel to the scintillation velocity one observes identical
fluctuations in both components, but with a delay corresponding to
the angular separation of the components relative to the scale of
the scintillation pattern.  If the velocity is orthogonal, one
sees a loss of correlation, with a large decorrelation if the
angular separation of the components is greater than the relevant
scintillation decorrelation length divided by distance to screen
(i.e.\ $\theta_{\rm F}$ in the weak scattering regime).

However, real sources are more complicated than two blobs.  VLBI
polarimetric imaging (e.g.\ Gabuzda, Pushkarev, \& Cawthorne 2000)
shows a great richness of detail and structure in the linear
polarisation of milliarcsecond components in AGN.  If IDV sources
exhibit the same richness of structure on microarcsecond scales
that they do on VLBI scales it would be surprising that the
polarisation lightcurves of IDV sources bear any resemblance at
all to the fluctuations in total intensity! Indeed, many IDV
sources show rich structure in polarisation lightcurves relative
to the total intensity.

This suggests that attempts to model source variability in terms of a
finite number of distinct components are missing the point.  One instead
ought to be talking about measuring the power spectrum of the brightness
distribution in each of the four Stokes parameters.

It is possible to invert for the source structure in a rigorous
manner if the variability is due to scintillation.   Formally, the
power spectrum of the observed fluctuations is the product of the
scintillation power spectrum due to a point source, which is known
or can be determined, and the power spectrum of the source angular
brightness distribution; see Macquart \& Jauncey (2001 submitted) for more
detail, and Figure 2 for an intuitive explanation of the effect of
source structure on its lightcurve.  This technique would be
useful in testing the model advanced by Begelman for which one
expects small-scale structure in the CP.

\begin{figure}
\centering
\begin{center}
\centerline{\psfig{file=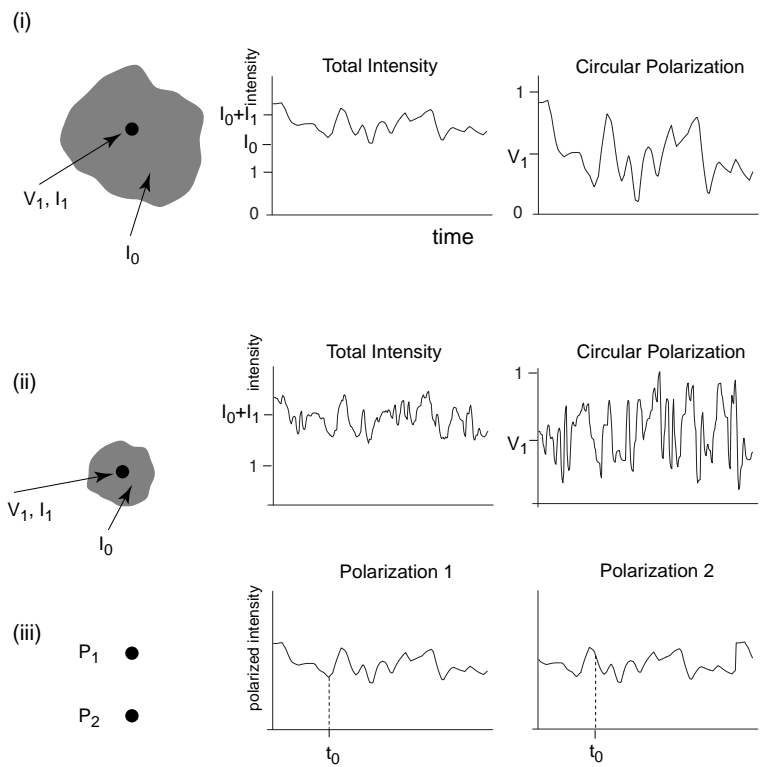,width=100mm}}
\end{center}
\caption[Extracting source structure]{Typical lightcurves for
common source configurations: (i) a polarised sub-source embedded
in a source too large to scintillate, (ii) two scintillating
sources, with the smaller component embedded in the larger, and
(iii) two distinct polarised components. }
\end{figure}

\section{Lessons from Other Sources}
\noindent{\bf Sgr~A*}

\noindent CP has been detected in Sgr~A* from 1.4 to 8.6 GHz with
ATCA and the VLA (Bower, Falcke, \& Backer 2000; Sault \& Macquart
1999). The mean degree of CP is 0.3\% but the CP varies on a
timescale of $\sim 10$~days, much faster than the monthly to
yearly variability observed in the total flux density at these
frequencies. The origin of the variability is not well
constrained; it could be intrinsic or due to scintillation of a
compact component in the source.

The spectrum of the mean degree of CP is flat from 1.4 to 8.6~GHz
but the significance of this is complicated by the fact that one
does not know the degree of CP in the polarised component.  It
appears that only a small portion of the source is polarised.
Changes in the CP are accompanied by only small changes in the
total intensity.  If the variability in the CP is indeed
associated with the variability of a specific component of the
source, the 1.4 and 2.4~GHz ATCA data suggest the presence of a
subcomponent with an intrinsic CP of a few percent.

\noindent {\bf Galactic microquasars}

\noindent CP is observed in several Galactic microquasars.  CP was
first detected in SS433 by Fender \et\ (2000) but the data were
insufficient to determine the origin of the CP. While the total
intensity emission and linearly polarised emission were
unresolved, the CP was not. It was suggested that the CP
originated in the inner parts of the system and that the actual
polarisation of these regions was as high as 10\%.

CP has recently been discovered in GRO~1655$-$40 from ATCA archival data
of the August 1994 outburst 
(J.-P. Macquart et al., in preparation).  The time evolution of
the CP was observed at frequencies from 1.4 to 9~GHz, and shows
the CP changing sign during the early stages of the outburst,
reaching a maximum and decaying over a period comparable to the
decay of the total emission.

\section{Conclusions}
The advent of high precision circular polarimetry has enabled a
reassessment of the CP in AGN. However, most of the outstanding questions
twenty years ago regarding the origin of the CP in AGN still remain.

VLBI and variability (scintillation) measurements indicate the CP
in AGN is associated with the emission near their cores, and that
it occurs in very compact regions.  Future measurements of the CP
will therefore only be useful if the observations are capable of
resolving this polarised emission.

The best means of obtaining this resolution at the centimetre
wavelengths at which the CP is currently being detected is via
scintillation `imaging'.  ATCA data indicate that a large
proportion of IDV sources exhibit variable CP.  Scintillation data
can be used to extract information about the degree of the CP and
its location relative to the linearly polarised and unpolarised
emission.

High precision polarimetry over a larger range of frequencies
would be useful in constraining the origin of the circularly
polarised emission. However, its use may be limited.
Complications due to source inhomogeneity and structure limit the
degree to which one can use the spectral slope of the CP to
constrain its origin.


\section*{Acknowledgements} It is a pleasure to thank Don Melrose, Dave Rayner, and
Bob Sault for their advice and help.

\section*{References}





\reference Benford, G.\ 1984, ApJ, {282}, 154

\reference Benford, G.\ 1992, ApJ, {391}, L59

\reference Blandford, R.D., \& K\"onigl, A.\ 1979, ApJ, 232, 34

\reference Bower, G.C., Falcke, H., \& Backer, D.C.\ 1999, ApJ,
{523}, L29

\reference Fender, R., Rayner, D., Norris, R., Sault, R.J., \&
Pooley, G.\ 2000, ApJ, 530, L29

\reference Gabuzda, D.C., Pushkarev, A.B., \& Cawthorne, A.W.\
2000,  MNRAS, 319, 1109

\reference Homan, D.C., \& Wardle, J.F.C.\ 1999,  AJ, 118, 1942

\reference Homan, D.C., Attridge, J.M., \& Wardle, J.F.C.\ 2001,
ApJ, 556, 113

\reference Jones, T.W., \& O'Dell, S.L.\ 1977, ApJ, {215}, 236

\reference Kennett, M.P., \& Melrose, D.B.\ 1998, PASA, {15}, 211

\reference Komesaroff, M.M., Roberts, J.A., Milne, D.K., Rayner,
P.T., \& Cooke, D.J.\ 1984,  MNRAS, {208}, 409

\reference Legg, M.P.C. \& Westfold, K.C.\ 1968, ApJ, {154}, 99

\reference Macquart, J.-P., \& Melrose, D.B.\ 2000a,
Phys.\,Rev.\,E, {62}, 4177

\reference Macquart, J.-P., \& Melrose, D.B.\ 2000b,  ApJ, {545},
798

\reference Macquart, J.-P., Kedziora-Chudczer, L., Rayner, D.P.,
\& Jauncey, D.L.\ 2000, ApJ, {538}, 623

\reference Manchester, R.N., Taylor, J.H., \& Huguenin, G.C.\
1975, ApJ, 196, 83

\reference Melrose, D.B.\ 1971, Ap\&SS, 12, 172

\reference Melrose, D.B.\ 1991, Ann.\,Rev.\,Astron.\,Astrophys., {29}, 31

\reference Pacholczyk, A.G.\ 1973, MNRAS, {163}, 29P

\reference Rayner, D.P., Norris, R.P., \& Sault, R.J.\ 2000,
MNRAS, 319, 484

\reference Saikia, D.J., \& Salter, C.J.\ 1988, Ann.\,Rev.\,Astron.\,Astrophys., 
{26}, 93

\reference Sault, R.J., \& Macquart, J.-P.\ 1999, ApJ, 526, 85L

\reference Sazonov, V.N.\ 1969,  Zh.\,Eksper Teor.\,Fiz., 56, 1074

\reference Weiler, K.W., \& de Pater, I.\ 1983, ApJ\,Supp, {52} 293

\reference Windsor, R.A. \& Kellog, P.J.\ 1974, ApJ, {190}, 167


\end{document}